# Slow spin relaxation in a highly polarized cooperative paramagnet

B. G. Ueland,[1] G. C. Lau,[2] R. J. Cava,[2] J. R. O'Brien,[3] and P. Schiffer[1*]

[1]*Department of Physics and Materials Research Institute, Pennsylvania State University, University Park PA 16802*

[2]*Department of Chemistry and Princeton Materials Institute, Princeton University, Princeton, NJ 08540*

[3]*Quantum Design, San Diego, CA 92121*

## Abstract

We report measurements of the ac susceptibility of the cooperative paramagnet $Tb_2Ti_2O_7$ in a strong magnetic field. Our data show the expected saturation maximum in $\chi(T)$ and also an unexpected low frequency dependence (< 1 Hz) of this peak, suggesting very slow spin relaxations are occurring. Measurements on samples diluted with nonmagnetic $Y^{3+}$ or $Lu^{3+}$ and complementary measurements on pure and diluted $Dy_2Ti_2O_7$ strongly suggest that the relaxation is associated with dipolar spin correlations, representing unusual cooperative behavior in a paramagnetic system.

[*]*schiffer@phys.psu.edu*

Geometrically frustrated magnets, in which the local spin-spin interactions compete with each other due to the lattice structure, have drawn a great deal of recent attention [1]. Such systems have been shown to display a variety of exotic behavior at low temperatures, including apparent glassiness in the presence of minimal disorder [2], spin-liquid like states, in which the spins continue to fluctuate even as $T \to 0$ [3,4,5], and "spin ice" -- a magnetic analog to the frustrated protonic state found in frozen water [6,7,8,9]. In particular, there has been extensive study of the effect of applied magnetic fields on these materials, since the application of a field restricts the available spin states and can lead to the development of long range order from a disordered state [3,10,11].

One frustrated system which has drawn particular interest is the rare-earth pyrochlore, $Tb_2Ti_2O_7$, in which the magnetic $Tb^{3+}$ cations form a sublattice of corner sharing tetrahedra, leading to frustration of their antiferromagnetic interactions. The $Tb^{3+}$ spins in this material show no signs of long range magnetic order down to temperatures below $T = 50$ mK, despite the development of nearest neighbor antiferromagnetic correlations below $T \approx 50$ K [4,12,13], and a mean field interaction strength indicated by $|\theta_W| \approx 18$ K [4,14,15,16,17,18]. This behavior indicates that $Tb_2Ti_2O_7$ is a cooperative paramagnet, or classical spin liquid [19], where the spins are correlated, yet fluctuate as $T \to 0$ [4]. The crystal field (CF) environment is somewhat analogous to that in the spin ice materials $Dy_2Ti_2O_7$ and $Ho_2Ti_2O_7$, in that a strong [111] anisotropy at $T \ll 20$ K has been predicted [17,20]. In contrast to spin ice, which has well defined Ising ground states with the lowest excited CF levels ~ 200-300 K removed, $Tb_2Ti_2O_7$ has a $J \approx 4$ ground state doublet and then another $J \approx 5$ CF doublet level about 20 K higher in energy [14,15,16,17,18]. While no long range ordering has been observed at ambient pressure, the antiferromagnetic spin interactions lead to a state exhibiting static long range order coexisting with dynamical states, at high pressure [21]. Additionally, zero field ac susceptibility measurements performed at dilution refrigerator temperatures have shown features characteristic of a glassy state below $T \approx 300$ mK [12,22,23].

Here, we report a study of the frequency-dependent magnetic susceptibility of $Tb_2Ti_2O_7$ in the presence of a strong applied magnetic field. We find evidence of unusually slow relaxation in the polarized state of this material at relatively high temperatures ($T > 20$ K), which appears to be a cooperative phenomenon. We observe similar behavior in another frustrated pyrochlore, the spin ice compound $Dy_2Ti_2O_7$, indicating that this field-induced behavior does not arise from the specific CF states of $Tb^{3+}$. Instead, we hypothesize that the frequency dependence may be a generic feature of highly polarized and strongly correlated paramagnetic systems, arising from the development of correlated regions coupled through dipolar interactions, akin to domains in a ferromagnet.

Polycrystalline samples of $(Tb_xM_{1-x})_2Ti_2O_7$ and $(Dy_xM_{1-x})_2Ti_2O_7$, where M = Lu or Y, and $1 \geq x \geq 0.01$, were prepared through standard solid state synthesis techniques and were determined to be single phase by x-ray diffraction. We used a Quantum Design MPMS SQUID magnetometer to measure the magnetization ($M$) as well as the low frequency ac susceptibility ($f < 10$ Hz). Higher frequency (10 – 10,000 Hz) ac susceptibility was measured using the ACMS option of a Quantum Design PPMS cryostat. All measurements were parallel to the direction of the applied dc magnetic field. For the ac measurements, data were independent of the magnitude of the excitation field between $H_{ac} = 0.5 - 10$ Oe. We also obtained the susceptibility from the magnetization data, $[dM/dH]_{dc}$, using measurements taken with increasing temperature over a range of static fields and then taking the derivative with respect to field at constant temperature. Note that these susceptibility measurements are not truly in the static dc limit, since the MPMS makes measurements at timescales of order $10^2$ seconds. There were no differences between zero-field-cooled (ZFC) and field-cooled (FC) data, even after cooling from $T > 100$ K, so we present ZFC results in all of the figures below, cooled from at least $T = 100$ K.

Curie-Weiss fits to the magnetization data [24] show effective magneton values equivalent to the accepted $p = 9.5 \pm 0.5$ for free $Tb^{3+}$ and $p = 10.5 \pm 0.5$ for free $Dy^{3+}$, for all samples. The Weiss temperatures, $\theta_W$, also compare well to those previously reported, where we find $\theta_W = -17.9 \pm 0.3$ K for $x = 1$ samples and $\theta_W = -7.4 \pm 0.4$ K for $x = 0.01$ $(Tb_xM_{1-x})_2Ti_2O_7$ samples [4,17]. Similar fits on the $(Dy_xM_{1-x})_2Ti_2O_7$ samples yielded $\theta_W \sim 1$ K, also consistent with previous reports [6].

In Fig. 1, we show the real and imaginary parts of the ac susceptibility, $\chi'(T)$ and $\chi''(T)$, of undiluted $Tb_2Ti_2O_7$ at various excitation frequencies and dc applied magnetic fields, from data taken in the PPMS ($f = 10$-$10,000$ Hz). At $H = 0$, $\chi'(T)$ closely follows the canonical paramagnetic behavior, i.e., monotonically rising with an increasing slope as the temperature is decreased, and the data can be fit with a combination of the Curie-Weiss formula and a Van Vleck term (a term for the susceptibility due to level transitions)[17,25]. By contrast, data taken in a field are more complex. We observe a peak at $T \sim 20$ K in $\chi'(T)$ for fields $H > 3$ T, which shifts to higher temperatures with increasing field. The approximately temperature independent susceptibility for $T \to 0$ in the largest fields comes from the Van Vleck terms, due to the closely spaced $Tb^{3+}$ CF levels [17,18,25,26]. The high field peak in $\chi'(T)$ is a consequence of single moment saturation: since the susceptibility associated with thermal spin fluctuations must approach zero both at high temperatures and as $T \to 0$ in the presence of a strong enough field. The imaginary part of the susceptibility $\chi''(T)$ shows a peak corresponding to the drop in $\chi'(T)$ as expected from the Kramers-Kronig relations. While one would expect this peak to be frequency independent on the time scale of our measurements, we observe a slight frequency dependence of the features in both $\chi'(T)$ and $\chi''(T)$, and the peak in the latter is clearly growing with decreasing frequency, even at $f = 10$ Hz. This frequency dependence at the single moment polarization peak in $\chi'(T)$ indicates the presence of unusually slow spin relaxation in this nominally paramagnetic material.

The bottom panel of Fig. 2 shows typical magnetization data used to obtain $[dM/dH]_{dc}$, and the top panel shows $H = 5$ T data for $[dM/dH]_{dc}$ and $\chi'(T)$, including both data from Fig. 1 and lower frequency ac data taken with the MPMS. The $[dM/dH]_{dc}$ data are strikingly different from the ac data, with the polarization peak occurring at much lower temperatures and with larger magnitude. The difference confirms the expectation from Fig. 1, that strong frequency dependence to the susceptibility exists at low frequencies. We also see that the lower frequency ac data progressively approach the $[dM/dH]_{dc}$ data as the frequency is decreased. It is notable that even for $f = 0.5$ Hz, the peak in $\chi'(T)$ is still only 80% the magnitude of the $[dM/dH]_{dc}$ peak, indicating that the system is still far from an isothermal susceptibility [27].

To further investigate the nature of this very slow spin relaxation, we studied samples of $(Tb_xM_{1-x})_2Ti_2O_7$, where M = Y or Lu, which are non-magnetic, replace the Tb ions, thus diluting the magnetic sublattice. In the top panels in Fig. 3, we plot the temperature dependence of $\chi'$ at $f$ = 10 Hz and $[dM/dH]_{dc}$ for various fields for both the pure sample as well as dilute samples with $x$ = 0.01. For both the Y and the Lu samples, the high dilution eliminates the low frequency dependence of the polarization peak, causing the differences in magnitude between $\chi'$ and $[dM/dH]_{dc}$ to essentially disappear.

We now consider possible explanations for the observed frequency dependence of the polarization peak, which implies that spin relaxation is occurring over very long time scales in this polarized paramagnet. The absence of frequency dependence at higher frequencies rules out simple thermal relaxation of individual spins, which would lead to an Arrhenius relation as was observed in $Dy_2Ti_2O_7$ at lower fields [9]. There is some qualitative resemblance to spin glass behavior, but the absence of frequency dependence at higher frequencies and the onset of the frequency dependence only in a strong field would appear to rule out a spin glass state as an explanation. Also, the absence of differences between ZFC and FC magnetization data in our temperature and field ranges appears to rule out a glass state. Another possibility is that the slow relaxation is associated with an unusual spin-phonon interaction that is sufficiently rare to cause an atypically slow relaxation rate. A change in the phonon spectrum associated with the mass differences between the $Tb^{3+}$ and the non-magnetic diluting ions could then change the phonon-mediated spin relaxation and would be responsible for the slow spin relaxation. This explanation does not seem plausible, however, since there is a difference of a factor of two between the masses of Lu and Y, with Lu near the mass of Tb, and thus any changes to the phonon spectrum should be drastically different for the two different diluting species. One could consider whether the observed phenomena are related to the onset of long range order, and recent measurements have indeed suggested the existence of magnetic long range order in $Tb_2Ti_2O_7$.[28] On the other hand, those observations are at much lower temperatures than where we observe the slow relaxation. The fact that the observed slow relaxation is coincident with the feature associated with single moment saturation (and is also seen in $Dy_2Ti_2O_7$, as discussed below) suggests that the origin lies within the highly polarized paramagnetic state.

One more possible explanation for our data is that there is an anomalous relaxation associated with the closely spaced CF level structure of the $Tb^{3+}$ in the presence of a strong dc field. To further investigate this possibility, we examined the diluted spin ice material $(Dy_xM_{1-x})_2Ti_2O_7$, which we have studied previously, but not in such strong magnetic fields. The data from these samples are given in the bottom half of Fig. 3, and they appear strikingly similar to those for $(Tb_xM_{1-x})_2Ti_2O_7$. In pure $Dy_2Ti_2O_7$, we again observe a substantial difference between $\chi'$ and $[dM/dH]_{dc}$ in a field, with the peak much larger and at lower temperatures in $[dM/dH]_{dc}$ than in $\chi'$. This difference disappears in samples highly diluted with either $Y^{3+}$ or $Lu^{3+}$, just as in the case of $(Tb_xM_{1-x})_2Ti_2O_7$. Since the CF level structures are substantially different for the $Tb^{3+}$ and $Dy^{3+}$ titanates, these data discount the possibility that the slow spin relaxation is a single ion effect associated with relaxation between CF levels.

While none of the above explanations appear plausible, the elimination of the frequency dependence of the data with dilution suggests that the implied slow spin relaxation originates with the development of spin correlations. A macroscopic effect from such correlations would be quite surprising for a paramagnetic material in which the spins do not freeze down to much lower temperatures. On the other hand, rare earth pyrochlores have large moments and long range dipolar interactions of order 1 K that are important in determining their low temperature properties [29,30], and previous studies have hypothesized a cooperative state at lower fields [31]. Since these systems are highly polarized in the dc fields we are considering (see Fig. 2), one explanation is that we are observing the formation of correlated domains of spins highly polarized along the field, with a few polarized in the opposite direction, due to the dipolar interactions of groups of polarized spins competing with the external field. There will be large energy barriers to switching these few oppositely polarized spins, and thus the relaxation between spin states will be very slow. Slow dynamics have indeed been observed in ferromagnetic cluster systems (albeit still faster than in our materials), and also have been associated with domain dynamics in ferromagnets [32]. Consistent with this interpretation, measurements on different shaped samples show that sample shape (which is related to demagnetization fields around the sample) changes the effect only qualitatively -- altering the magnitudes of the susceptibility peaks but not the temperatures at which they occurred.

Our observation that strong magnetic fields may induce slow spin relaxation contradicts normal intuition that a strong field would hasten relaxation to an equilibrium state [33]. The data suggest a new form of collective behavior in the rare earth titanates and further indicate the importance of understanding the effects of magnetic fields on frustrated magnets. While this relaxation may be possible in any strongly correlated paramagnetic system, we note that the effects are particularly important to geometrically frustrated systems, because they remain paramagnetic down to temperatures that are low, relative to the strengths of the spin-spin interactions.

**ACKNOWLEDGEMENTS**

We gratefully acknowledge helpful discussions and exchanges with B. S. Shastry, R. Moessner, M. J. P. Gingras, B. D. Gaulin, G. Aeppli, S. R. Dunsiger, J. S. Gardner, A. P. Ramirez, and T. F. Rosenbaum and support from NSF grants DMR-0101318 and DMR-0353610.

**Figure captions**

**FIG. 1.** The temperature dependence of the real and imaginary parts of the ac susceptibility ($\chi'(T)$ and $\chi''(T)$) of $Tb_2Ti_2O_7$ at different frequencies in various applied magnetic fields.

**FIG. 2.** The magnetic susceptibility of $Tb_2Ti_2O_7$ at $H = 5$ T. The top panel includes ac measurements from the MPMS at low frequency and from the PPMS at higher frequency as well as $[dM/dH]_{dc}$ from magnetization data, as described in the text. The bottom panel shows the measured magnetization $M(T)$ at various magnetic fields ($H$ increases from bottom to top). Note the high degree of polarization at the fields where slow relaxation is occurring.

**FIG. 3.** Comparison of the ac magnetic susceptibility measured at $f = 10$ Hz and $[dM/dH]_{dc}$ for both $(Tb_xM_{1-x})_2Ti_2O_7$ and $(Dy_xM_{1-x})_2Ti_2O_7$ as described in the text. Note that the difference between the two is almost eliminated in the highly dilute samples. The dip in the $H = 0$ ac data of the diluted Dy samples is a consequence of simple single-ion relaxation studied previously [9], which unlike the frequency dependence we observe, follows an Arrhenius law.

Fig. 1. Ueland *et al.*

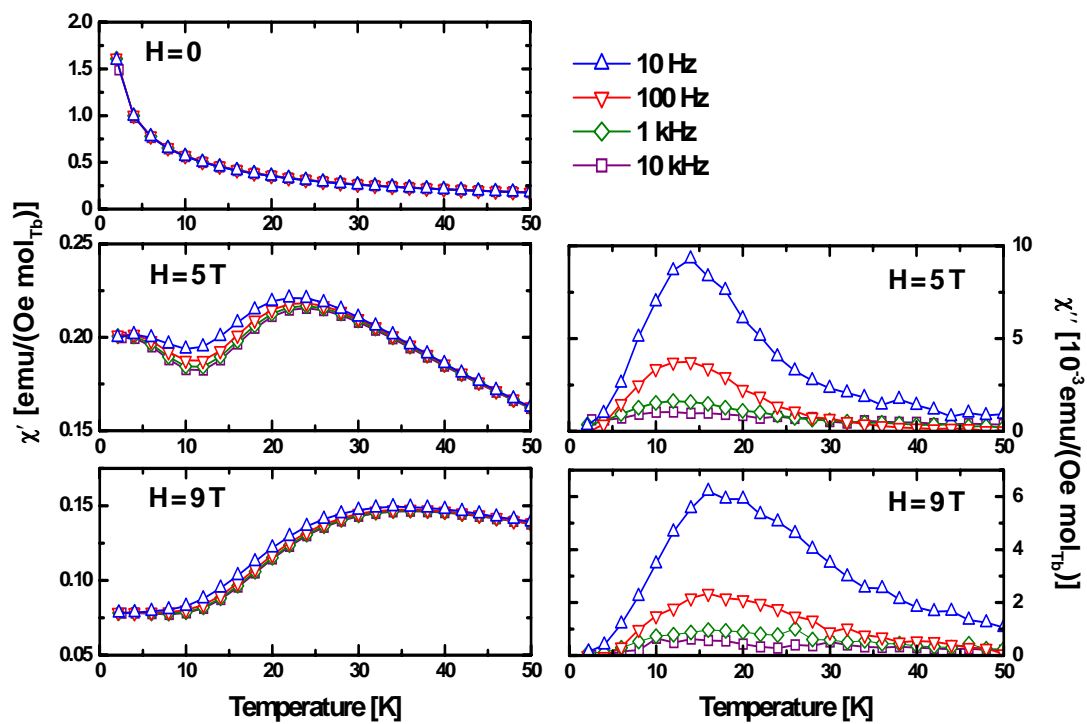

Fig. 2. Ueland *et al*.

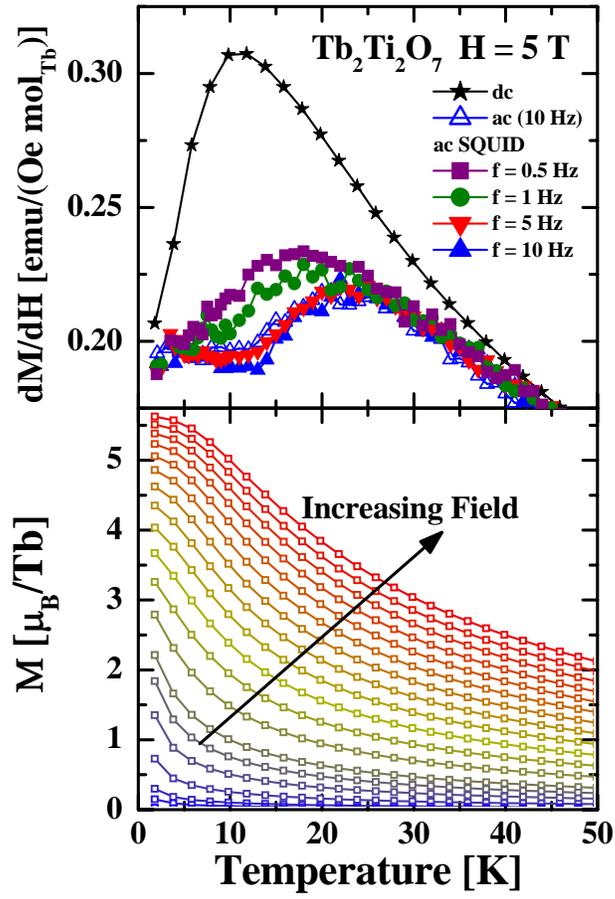

Fig. 3. Ueland *et al*.

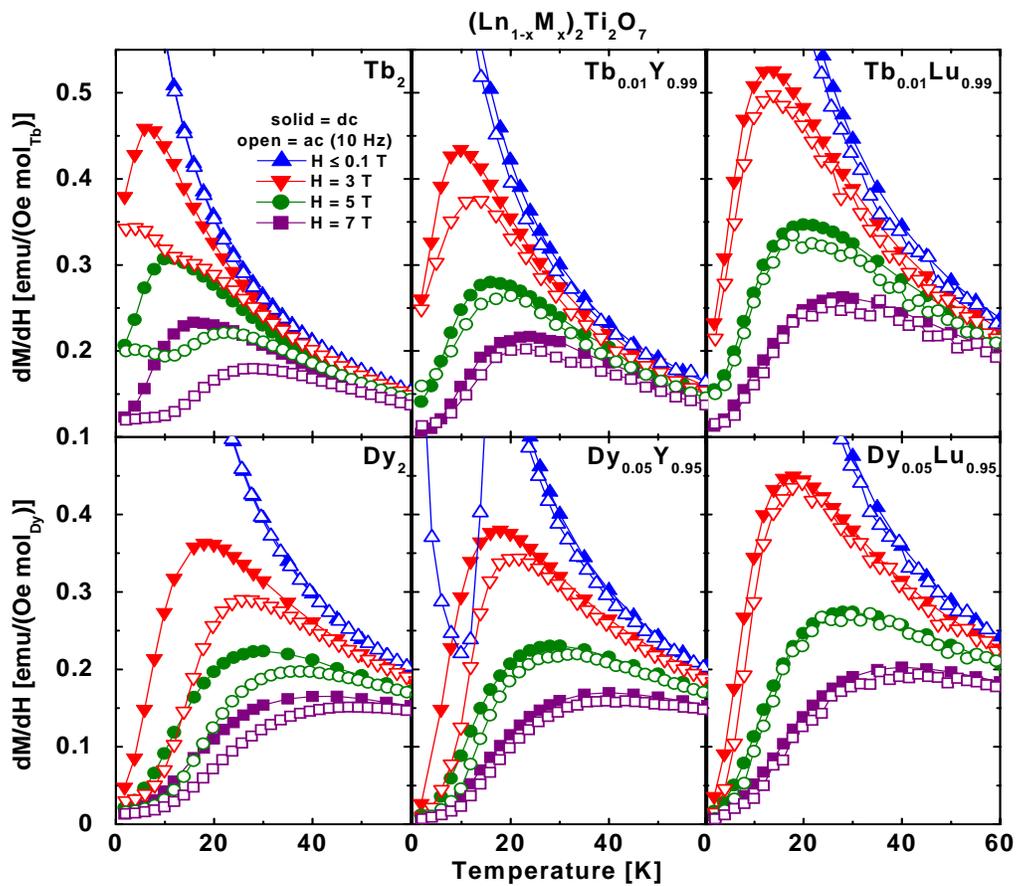